\documentclass[twocolumn,floatfix]{revtex4}
	\usepackage{amssymb}
	\usepackage{epsfig}
	\usepackage{amsmath}
	\usepackage{graphicx}
	
\begin{document}

\newcommand{\thresh}{{T_{\rm thresh}}}
\newcommand{\aopt}{{\alpha_{\rm opt}}}
\newcommand{\arew}{{\alpha_{\rm rewired}}}


	\title{How Do Networks Become Navigable?}
	\author{Aaron Clauset$^*$ and Cristopher Moore$^{\dagger,*}$}
	\affiliation{$^*$Computer Science Department and $^\dagger$Department of Physics and Astronomy, University of New Mexico, Albuquerque NM 87131 \\ 
	{\tt \{aaron,moore\}@cs.unm.edu}}
	\date{\today}

	\begin{abstract}
	Networks created and maintained by social processes, such as the human friendship network and the World Wide Web, appear to exhibit the property of  \emph{navigability}: namely, not only do short paths exist between any pair of nodes, but such paths can easily be found 
	using only local information.  It has been shown that for networks with an underlying metric,  algorithms using only local information perform extremely well if there is a power-law distribution of link lengths.  However, it is not clear why or how real networks might develop this distribution.  In this paper we define a decentralized ``rewiring'' process, inspired by surfers on the Web, in which each surfer attempts to travel from their home page to a random destination, and updates the outgoing link from their home page if this journey takes too long.  We show that this process does indeed cause the link length distribution to converge to a power law, achieving a routing time of $O(\log^2 n)$ on networks of size $n$.  
	We also study finite-size effects on the optimal exponent, and show that it converges polylogarithmically slowly as the lattice size goes to infinity.
	\end{abstract}
	\maketitle

	\section{Introduction}

	It is well-known that a wide variety of social networks have the so-called {\em small world} property; networks of size $n$ have diameter $O(\log n)$, meaning that between any two nodes there exists a path of size $O(\log n)$.
	This property is also shared by sparse random graphs; however, random graphs lack several other important properties of real-world networks, such as clustering and heavy-tailed degree distributions (see~\cite{review-structurefunction} for a review).
	
	Another property of social networks which distinguishes them from random graphs, and which has received somewhat less attention, is their {\em navigability}.  That is, not only do short paths exist, but it is easy to find them using only local information.  
	The pioneering work of Milgram in the 1960s \cite{basic-milgram} showed that people can, at least some of the time, find short paths to other distant people by pursuing the greedy strategy where they pass the message to whichever of their own acquaintances they feel is ``closest'' to the target recipient.  This experiment has recently been repeated using e-mail networks~\cite{email}.  
	In a random graph, although a short path exists, a local algorithm must be lucky to find it as it can do little better than a random walk on the network.
	
	Kleinberg~\cite{basic-kleinberg,kleinbergnature} addressed the issue of navigability by considering a small-world model consisting of a $d$-dimensional lattice with long-range links added to it. Unlike the Watts-Strogatz model~\cite{basic-collectiveswndynamics} in which the long-range links are uniformly distributed, in the Kleinberg model pairs of nodes a distance $\ell$ apart are connected with some probability $P(\ell)$, where $P(\ell) \sim \ell^{-\alpha}$ and the number of outgoing links per node is fixed.
	While a finite-dimensional lattice is obviously a gross oversimplification of the social spaces in which we live, or the conceptual spaces in which one Web page seems ``closer'' to another, this model captures the essential features of the navigation problem: how can we minimize the routing time 
	given local information about some metric?
	
	Kleinberg studied the performance of the greedy algorithm, in which each node passes the message to whichever of its neighbors, either local or long-range, that is closest to the destination on the underlying lattice.  He showed that if $\alpha=d$  this algorithm achieves a routing time of $O(\log^2 n)$ on lattices of size $n$.  However, if $\alpha \ne d$, the routing time grows as $O(n^\beta)$ for some $\beta > 0$.  
	Note that if when $\alpha=d$, we integrate over spheres of radius $\ell$, we find that the distribution of link lengths is $f(\ell) \sim \ell^{-1}$, with a cutoff when $\ell$ reaches the system size; this distribution provides the right mix of long-, medium-, and short-range links for the greedy algorithm to quickly ``zero in'' on its destination.  Similar results have been obtained for the case where the underlying graph is a tree, representing a hierarchical structure, or an overlapping set of trees representing  multiple affiliations in society~\cite{kleinberg-trees,wdn}.

	While Kleinberg's work largely characterizes the {\em static} properties a network needs to have to be navigable, it leaves open the question of how or why networks might evolve these properties.  Indeed, the topology of a social network is constantly being modified by its  members.  If, whenever these members find it difficult to search the network, they modify their own connections in an effort to make future searches easier, this dynamical process should make the network more navigable over time.  
		
	In this paper, we consider a dynamic network model inspired by surfers on the Web, each of whom controls the outgoing links from their home page.  We start at a source node $x$, and choose a random destination node $y$.  Based on the (metric) distance between $x$ and $y$, we set a threshold $\thresh$ on the number of (topological) steps we feel the journey should take.  
	If after $\thresh$ steps we have not reached our destination, we stop searching, and rewire $x$'s long-range link to the place where we gave up.  In the human-friendship network, this might correspond to gaining new acquaintances in the course of a search; on the Web, it corresponds to creating a bookmark/linking to the relevant pages, which other surfers can then use~\cite{note}.
	
	Our main result is that this process does indeed cause the distribution of link lengths to converge to a power law $f(\ell) \sim \ell^{-\arew}$.  The precise value of $\arew$ varies with lattice size, and differs somewhat from Kleinberg's prediction $\alpha = d$.  We believe this is due to finite-size effects; to support this belief, we directly construct networks with power-law link length distributions (as opposed to rewiring them), measure the exponent $\aopt$ that minimizes the routing time, and find that it converges rather slowly to $d$ as $n \to \infty$, roughly as $d - O(1/\log^2 n)$.   
	However, even though the exponents differ somewhat, our rewiring process produces networks whose routing times match or improve those of Kleinberg's optimum.
	
	Other types of dynamical models have been studied (e.g.~\cite{rewiring-euclidspace, rewiring-hamilton, rewiring-ffp, growing-attach, growing-walk}).  However, the model discussed here appears to be the first to optimize for navigability.
	
	\section{The Rewiring Process and Results}
	
	One restriction of Kleinberg's model of a social space consists of a $d$-dimensional lattice of size $n$, where each node $u$ is connected to its $2d$ nearest neighbors and has a single long-range link to a node $v$, chosen with \mbox{probability} $P(\ell) \sim \ell^{-\alpha}$ where $\ell=|u-v|$ is the Manhattan distance (the $L_1$ norm).  Such networks are denoted $G(n,d,1,1,\alpha)$ in~\cite{basic-kleinberg}; here the two $1$s indicate the radius of the local \mbox{connections} and the number of long-range links per node.  Our goal is to show that our rewiring process causes networks with a range of initial link length distributions to converge to this form, where $\alpha = \arew$ and is close to Kleinberg's prediction $\alpha = d$.  We report here on \mbox{experiments} where $d=1$, for which we can feasibly study networks of size up to $n=10^7$; our results for $d=2$ are qualitatively similar.
	
	In each round of the rewiring process, we choose the source node $x$ uniformly, and choose the destination $y$ according to a {\em demand distribution} $Q(x,y)$.  One method of selection is to choose a distance $\ell$ according to a distribution $Q(\ell)$, and then choose $y$ randomly from among the nodes that satisfy $|x-y| = \ell$. Here, we take $Q(\ell)$ to be uniform, which for $d=1$ means that $y$ is simply a uniformly random node; we comment in the Conclusions on the effect of other demand distributions.  

	Starting at $x$, the greedy algorithm produces a path $x=x_0,x_1,x_2,\ldots,x_T=y$ where $T$ is the routing time, and where for each $0 \le i < T$, 
$x_{i+1}$ is the outgoing neighbor of $x_i$ that minimizes $|x_{i+1}-y|$ (with ties broken randomly).  The rewiring process works as follows: if $T \geq \thresh$, i.e.,\ if the greedy algorithm reaches some $x_i$ with $i = \thresh$, then we discontinue the search and change $x$'s outgoing link to point to $x_i$.
	
	In our experiments, we choose the threshold $\thresh$ uniformly at random from the interval $[1,\ell]$.  Such a naive selection avoids making assumptions about the network's size, performance or topology.  In particular, it avoids assuming the $O(\log^2 n)$ routing time which both the rewired networks and Kleinberg's networks achieve.
	
	For our initial conditions, the link length distributions have exponent $\alpha_0$ ranging from $0$, a uniform distribution as in the Watts-Strogatz model, to $\infty$, where all the ``long-range links'' are simply self-loops.  We run $\tau$ rounds per node of the rewiring process; that is, $\tau n$ times we choose nodes $x,y$, run the greedy algorithm from $x$ to $y$ and rewire if $T\geq\thresh$. We then compare the final link length distribution to Kleinberg's optimum $f(\ell) \sim \ell^{-1}$.

\begin{figure} 
\includegraphics[scale=0.39]{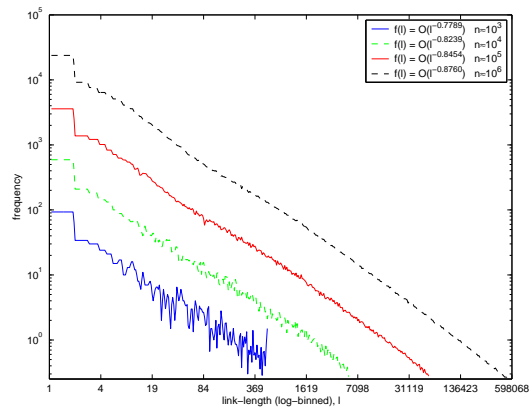}
\caption{Link length distributions with power-law fits for networks of size $n=\{10^{3},10^{4},10^{5},10^{6}\}$ after running the rewiring process until $T_{\rm rewired}\leq 1.01\cdot T_{\rm opt}$.  The initial graphs had $\alpha_0 = \infty$, i.e.,\ only self-loops.}
\label{fig:logbinned}
\end{figure}

\begin{figure} 
\includegraphics[scale=0.39]{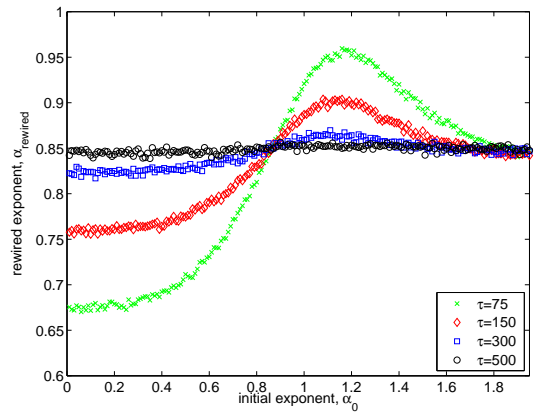}
\caption{Convergence of $\arew$ for a range of initial link length exponents $\alpha_0$, where $n=10^5$.  At $\tau$ increases, $\arew$ converges to $\arew \approx 0.85$ regardless of the value of $\alpha_0$.}
\label{fig:variousa} 
\end{figure}

	Figure~\ref{fig:logbinned} illustrates the resulting distributions for networks of size $n=\{10^{3},10^{4},10^5,10^{6}\}$ and $\alpha_0 = \infty$.  We see that after $\tau n$ rounds (we describe our choice of $\tau$ in the next section) the rewiring process
	 has built a power-law distribution of link lengths, $f(\ell) \sim \ell^{-\arew}$ with $\arew \approx \{0.78, 0.82, 0.85, 0.88\}$ respectively. 
	Figure~\ref{fig:variousa} shows that a range of initial distributions converge to the same final distribution.  Starting with initial exponents $\alpha_0$ ranging from $0$ to $2$, the exponent of the rewired link length distributions all converge to $\arew\approx0.85$ (here $n=10^{5}$) as the rewiring continues.

	\section{Finite-size effects}

	While our rewiring process produces a power-law link length distribution, the exponent is noticeably different from the value $\alpha = 1$ (more generally, $\alpha = d$) which Kleinberg proved in the limit $n \to \infty$.  Notably, Figure~\ref{fig:variousa} shows that networks with an initial exponent $\alpha_0=1$ actually move away from this value as they are rewired.
	
	This deflection appears to be due to finite-size effects, and in fact the exponent $\aopt$ that minimizes the routing time on finite lattices turns out to be rather different from Kleinberg's value even at reasonably large $n$.  We examined system sizes over six orders of magnitude, $10^{1} \leq n \leq 10^{7}$.  For each size, we constructed networks with $\alpha$ ranging from $0$ to $1$, and measured the mean routing time $T(\alpha)$ over $50n$ trials for each $\alpha$.  We estimated $\aopt$ as the minimum of the best quadratic fit of $T(\alpha)$.
	
\begin{figure} 
\includegraphics[scale=0.39]{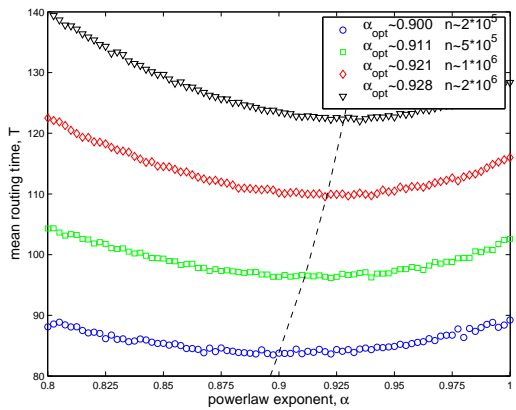}
\caption{\label{fig:static-egfit} Mean routing time $T(\alpha)$ for several network sizes. The dashed line shows the path $\aopt$, defined as the minimum of the best quadratic fit, follows as $n$ increases.}
\end{figure}

\begin{figure} 
\includegraphics[scale=0.39]{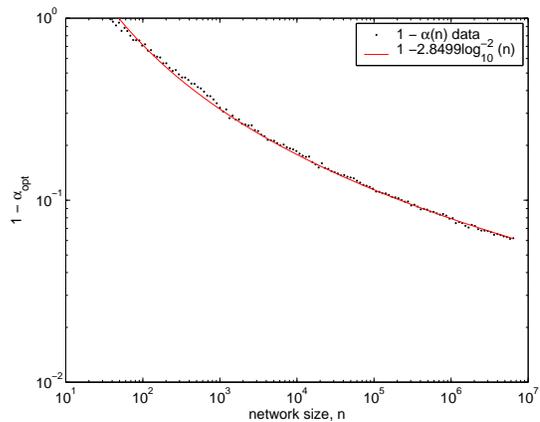}
\caption{\label{fig:static-rminasymptote} Finite-size dependence of $\aopt$ for $n$ up to $10^7$.  The solid line shows a fit to $1-\aopt \sim 1/\log^2 n$, showing that $\aopt$ converges polylogarithmically slowly to $d$ as $n \to \infty$.}
\end{figure}

	Figure~\ref{fig:static-egfit} illustrates $T(\alpha)$ for several values of $n$ and clearly shows that $\aopt < 1$. The dashed line illustrates $\aopt$'s slow approach to $1$ as $n$ increases.  Figure~\ref{fig:static-rminasymptote} shows the dependence of $\aopt$ on $n$; it is fit rather closely by the form $\aopt = 1 - A/\log^2 n$ for $A \approx 2.85$.  The fact that $\aopt$ converges polylogarithmically, as opposed to polynomially, indicates that finite-size effects are quite severe.
	
\begin{figure} 
\includegraphics[scale=0.39]{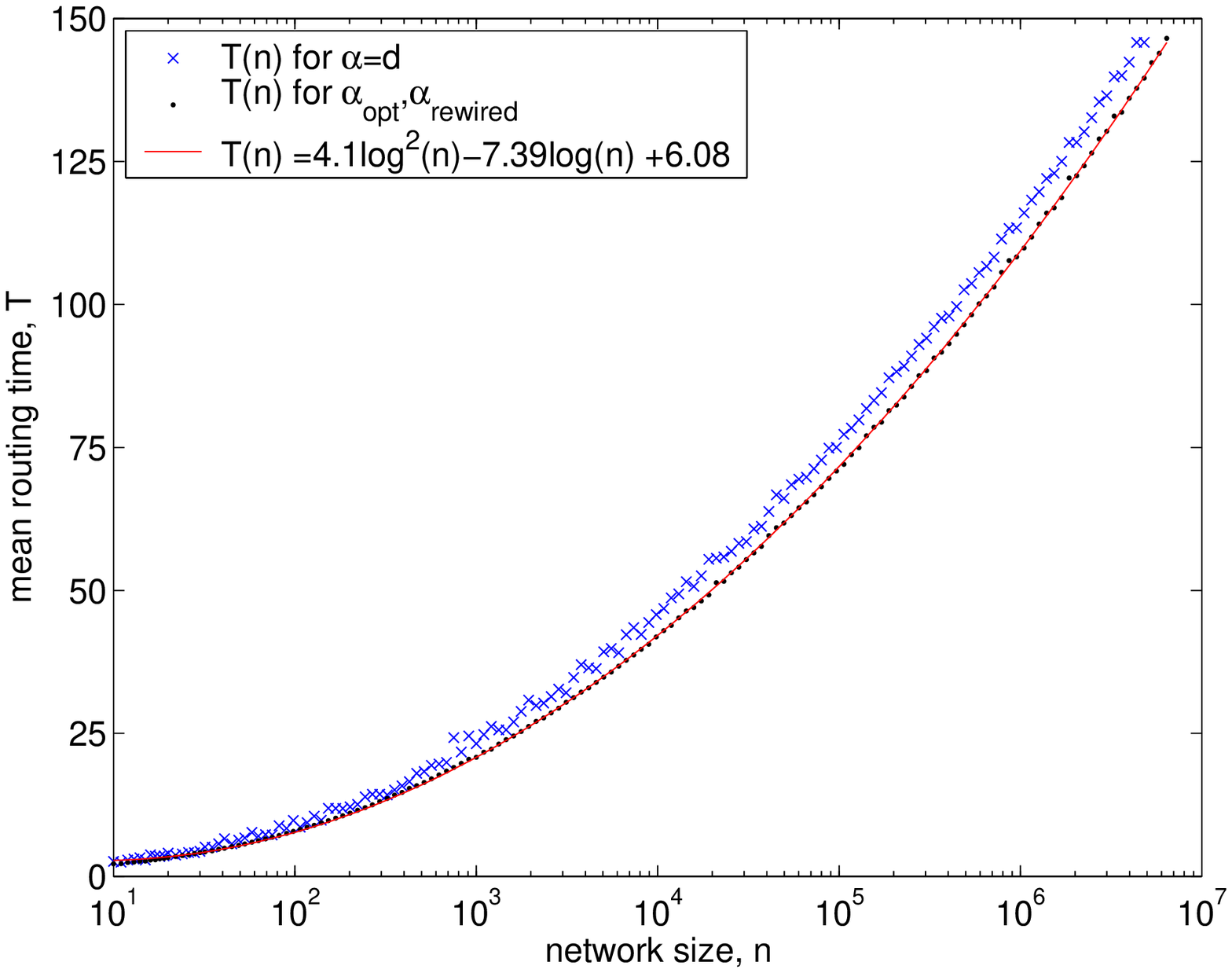}
\caption{\label{fig:static-meanrouting} Mean routing time $T$ for networks with $\alpha = \{d, \aopt,\arew\}$.  Shown is a fit to $\alpha=\arew$ of the form $T \sim \log^2 n$.}
\end{figure}

However, while $\arew$ differs from $d$ (and from $\aopt$) our rewired networks achieve a mean routing time equivalent to networks with exponent $\aopt$, and better than those with exponent $d$.   Fig.~\ref{fig:static-meanrouting} compares the mean routing times for the three types of networks.  All three are closely fit with a curve of the form $T(n) \sim \log^2 n$, just as Kleinberg's analysis predicts~\cite{basic-kleinberg,kleinbergnature}.

\begin{figure} 
\includegraphics[scale=0.39]{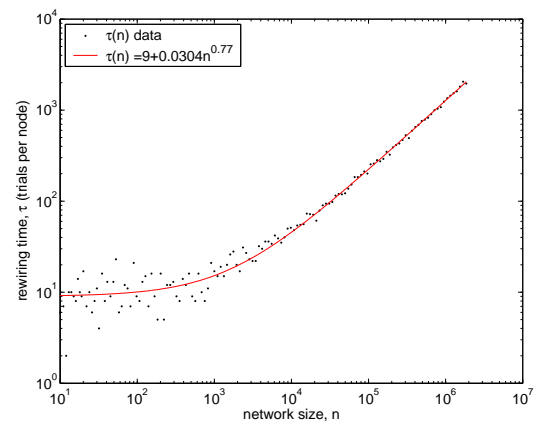}
\caption{The rewiring time $\tau(n)$, i.e.,\ the number of rounds per node the rewiring process needs to achieve a routing time $T_{{\rm rewired}}\leq 1.01\cdot T_{{\rm opt}}$, within $1\%$ of that of a network with exponent $\aopt$.  The solid line is a fit of the form $\tau \sim n^{0.77}+C$.}
\label{fig:relaxTime} 
\end{figure}

Of course, the routing time of the network depends on the number of rewiring rounds. In Figure~\ref{fig:relaxTime}, we show the number of rounds per node $\tau$ required to achieve a mean routing time $T_{\rm rewired}$ which is only $1\%$ greater than $T_{\rm opt}$, i.e.,\ the routing time of a network with exponent $\aopt$.  This {\em rewiring time} grows as $\tau \sim n^{0.77}$ rounds per node, or $n^{1.77}$ rounds total.

	\section{Discussion and Conclusions}
	
	Kleinberg~\cite{basic-kleinberg,kleinbergnature} explored the navigability of small-world networks built on an underlying $d$-dimensional space. In the limit of infinite size, he found the mean routing time in minimized by a power-law \mbox{distribution} of link lengths with exponent $\alpha=d$ \cite{note2}.  Here, we have explained how \mbox{social} \mbox{networks} might develop such a distribution over time, by \mbox{introducing} a \mbox{simple,} decentralized rewiring process that relies only on local information (and is even ignorant of the size of the network).

	This process has a natural interpretation: the topology of a social network is constantly being modified by its members, who update their personal connections as they explore and navigate the network.  If a \mbox{member} becomes frustrated because the journey to a destination takes too long, they can be expected to change their \mbox{connections} to make similar journeys more quickly in the future.  

	Our results show that rewiring causes a wide range of initial topologies to converge to a power-law distribution of link lengths, very similar to Kleinberg's. 
	The exponent $\arew$ we obtain differs significantly from $d$, its optimal value on infinite lattices.  We attribute this deflection to finite size effects which cause the optimal exponent $\aopt$ to converge polylogarithmically as $n \to \infty$.  However, the rewired network achieves the same mean routing time as power-law networks with exponent $\aopt$, and better routing times than those with $\alpha = d$.   Specifically, the mean routing time as a function of system size is $O(\log^2 n)$ as predicted by Kleinberg's analysis.  The number of rounds of the rewiring process needed to achieve this routing time grows as a low-degree polynomial of $n$. 
	
	In addition to creating and maintaining a power-law distribution of link lengths, we believe this rewiring process to be {\em adaptive}.  For instance, we conjecture that if new nodes are added to the network, or if certain nodes or links are removed, it will dynamically optimize for these new situations.  We also believe, based on preliminary results, that if the demand distribution $Q(x,y)$ is not uniform, e.g.\ if certain destinations are more popular than others, or if the source and the destination are correlated (both of which are true in any real network), that it will optimize routing times for the source-destination pairs that appear more frequently.  These adaptive properties would be particular useful in networks where nodes are constantly coming on- and off-line and where the demand of each destination rises and falls over time, such as live peer-to-peer networks, distributed sensor networks or massively parallel computers.
	
	Our discussion contemplates a ``social space'' consisting of a finite-dimensional lattice, an obviously poor model for the complex social spaces we routinely navigate.  An interesting study would be an analogous rewiring process for networks whose underlying structure is hierarchical~\cite{kleinberg-trees}, involves multiple group affiliations~\cite{wdn} or is otherwise structured, as in the peer-to-peer network Freenet~\cite{freenet} with the modifications described in~\cite{freenetrouting}.  We leave these as directions for future work.
	


\begin{acknowledgments}
The authors are grateful to Mark Newman and Jon Kleinberg for helpful discussions.  This work was funded by NSF grant PHY-0200909, and the Los Alamos National Laboratory.
\end{acknowledgments}



\begin{thebibliography}{2}
\bibitem{note} Note that our model conflates links and bookmarks; a surfer's bookmarks are located at her home node and may be utilized in others' navigation.
\bibitem{note2} In \cite{kleinbergnature}, Kleinberg noted that for a $n=20,000$, $d=2$ graph, the optimal value of $\alpha$ was slightly below $d$.
\end{thebibliography}
\end{document}